\pdfoutput=1
\documentclass[prl,preprint,aps,onecolumn,groupedaddress,showpacs,amsmath,amssymb]{revtex4-1}
\usepackage{graphicx}
\usepackage{amssymb}
\usepackage{ifpdf}
\usepackage{dcolumn}
\usepackage{bm}

\newcommand{\mutilde}{\tilde{\mu}}

\begin{document}

\title{Anomalous thermoelectric transport in two-dimensional Bose gas}

\author{Eric L. Hazlett$^1$, Li-Chung Ha$^1$,  and Cheng Chin$^{1,2}$}
\address{$^1$The James Franck Institute and Department of Physics, University of Chicago, Chicago, IL 60637, USA\\
$^2$The Enrico Fermi Institute, University of Chicago, Chicago, Illinois 60637, USA}
\date{\today}

\maketitle
\textbf{In condensed matter physics, transport measurements are essential not only for the characterization of  materials, but also to discern between quantum phases and identify new ones.  The extension of these measurements into atomic quantum gases is emerging \cite{EsslingerTransportScience,EsslingerNatureTransport,TunableSFJunction,BECinRing,CLHSlowTransport} and will expand the scope of quantum simulation \cite{FeynmanSimulator,QuantumSimulator} and atomtronics \cite{Atomtronic}. To push this frontier, we demonstrate an innovative approach to extract transport properties from the time-resolved redistribution of the particles and energy of a trapped atomic gas.  Based on the two-dimensional (2D) Bose gas subject to weak three-body recombination we find clear evidence of both conductive and thermoelectric currents. We then identify the contributions to the currents from thermoelectric forces and determine the Seebeck coefficient (a.k.a. thermopower) and Lorenz number, both showing anomalous behavior in the fluctuation and superfluid regimes. Our results call for further exploration of the transport properties, particularly thermoelectric properties, of atomic quantum gases.}\begin{figure}[t]
\includegraphics[width=6in]{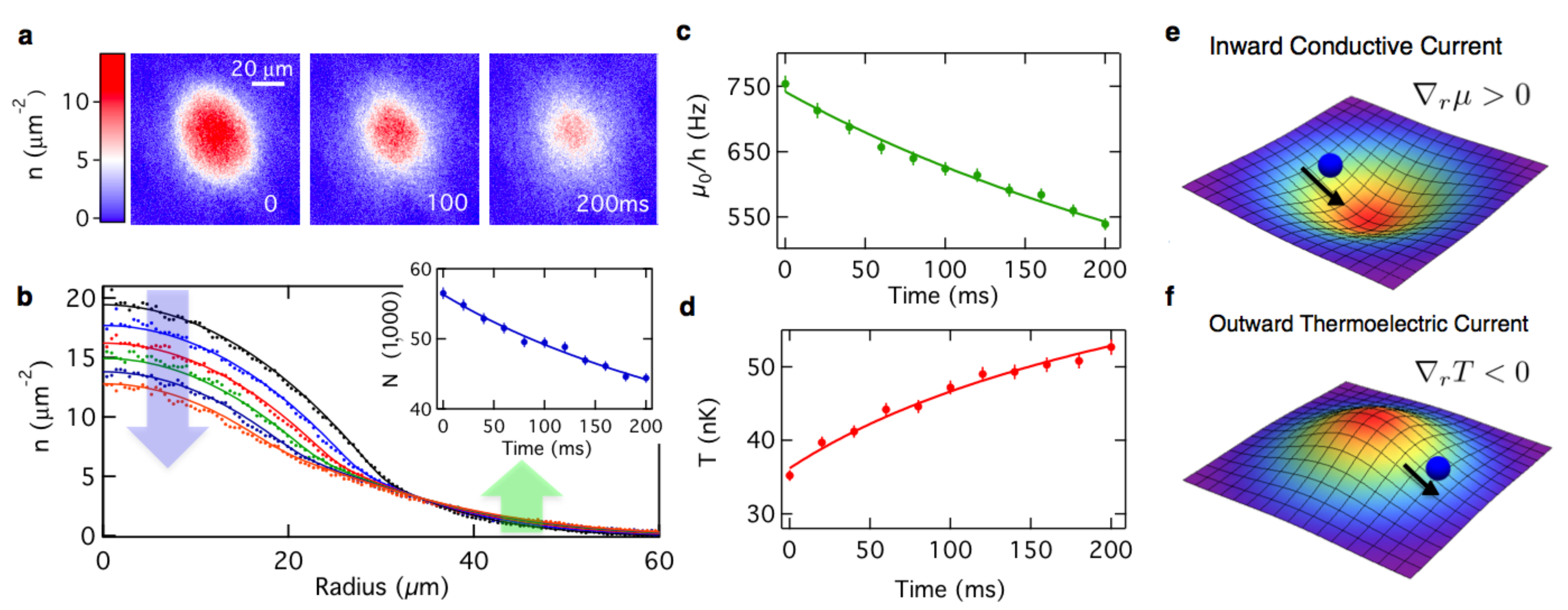}
\caption{\begin{bf}Evolution of 2D Bose gas driven by slow three-body recombination loss. \end{bf}  $\mathbf{a}$, $\mathit{in}$ $\mathit{situ}$ images of 2D Bose gases with scattering length $\mathit{a}=$ 22~nm at various hold times.  $\mathbf{b}$, Radial density profiles $n(r,t)$ (solid circles, 0$\sim$200 ms from top to bottom in 40~ms steps).  The solid lines are the parameterized density profiles based on the equation of state (see text and supplemental material). The blue and green arrows show regions where the density is decreasing and increasing, respectively. Inset shows the evolution of the total atom number $N$ and the fit. $\mathbf{c,d}$, The extracted chemical potential at the trap center $\mu_0$, the temperature $T$ and the fits. $\mathbf{e,f}$, Illustrations of chemical potential and temperature gradients which drive the particle flow inward (conductive current) and outward (thermoelectric current) in the density tail, respectively. }
\label{fig1}
\end{figure}

In transport theory, fluxes of extensive quantities are driven by thermodynamic forces.  For small deviations from equilibrium linear response theory applies, and particle flow, for example, is associated with a sum of a conductive current driven by the chemical potential gradient and a thermoelectric current driven by the temperature gradient. For an isolated system, transport theory relates the particle flux $j_p(r,t)$ and energy flux $j_e(r,t)$ to the thermodynamic conjugate forces $f_p=-\nabla(\mu/T)$ and $f_e=\nabla(1/T)$ by a linear equation \cite{MahanManyParticle}:

\begin{equation}
\begin{aligned}
&\left(\begin{array}{c} j_p\\ j_e \\ \end{array}\right)=\hat{\sigma}\left(\begin{array}{c} f_p\\ f_e \\ \end{array}\right),\\
\end{aligned}
\label{Fluxeqn}
\end{equation}
\noindent where $\mu$ is the chemical potential, $T$ is the temperature, and the kinetic matrix $\hat{\sigma}$ depends on the state of the matter and can be expressed in terms of the conductivity $\sigma$, the Peltier coefficient $\mathcal{P}$, the Seebeck coefficient $\mathcal{S}$, the thermal conductivity $K$, and the Lorenz number $\mathcal{L}=K/\sigma T$  as \cite{MahanManyParticle}
\begin{equation}
\hat{\sigma} = T\sigma\left(\begin{array}{cc}1 & T\mathcal{S}+\mu \\ \mathcal{P}+\mu &\;\;\; T^2\mathcal{L}+(T\mathcal{S}+\mu)( \mathcal{P}+\mu) \\ \end{array} \right).
\label{Fluxeqn1}
\end{equation}
\noindent  The Onsager relation suggests the kinetic matrix be symmetric $\hat{\sigma}^\intercal=\hat{\sigma}$, which gives $\mathcal{P}=T\mathcal{S}$\cite{OnsagerACS}.

\begin{figure}[t]
\includegraphics[width=3.4 in]{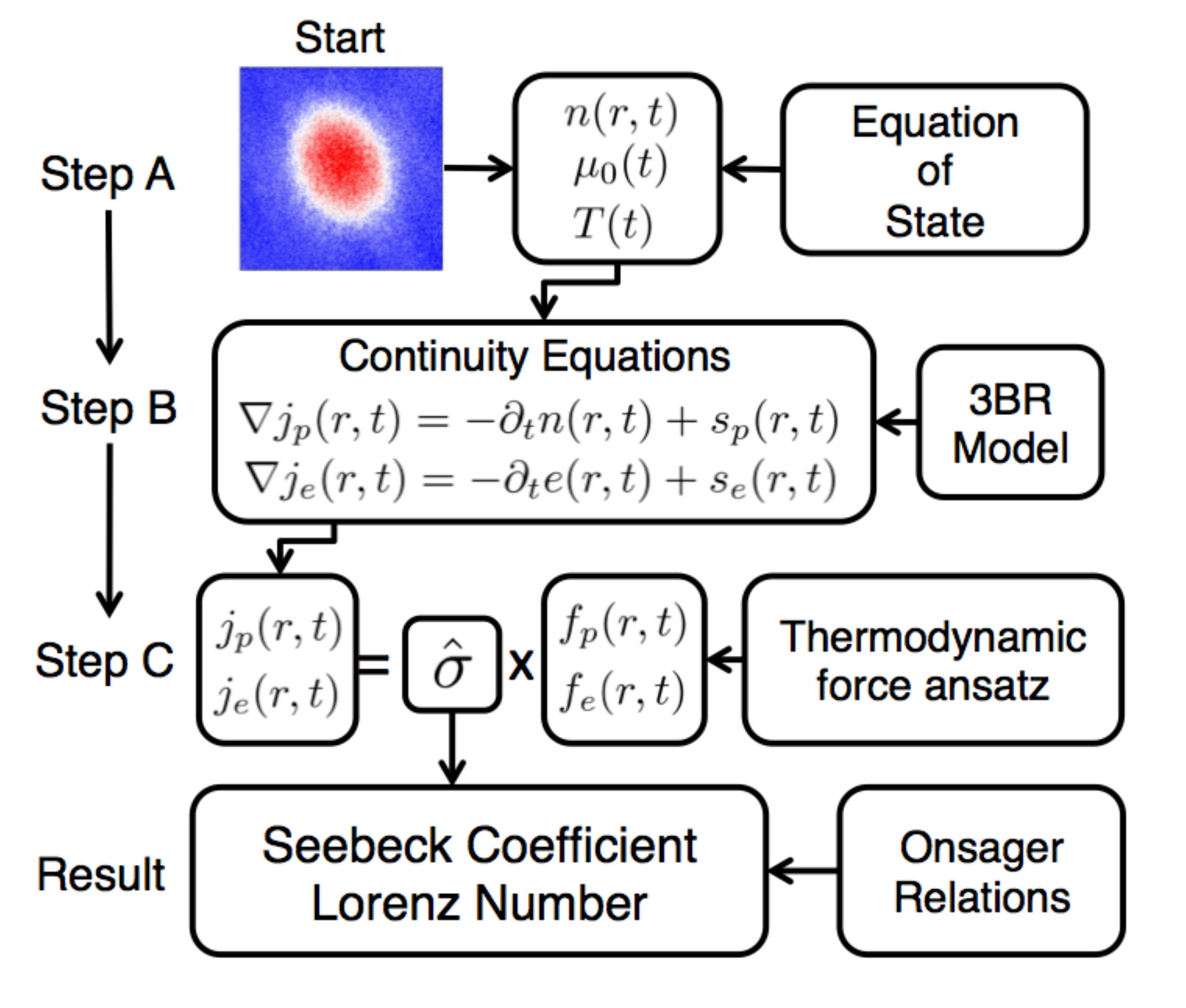}
\caption{\begin{bf}Method to extract transport properties from $\mathbf{\mathit{in~situ}}$ images \end{bf} Step A: We determine $\mu_0(t)$ and $T(t)$ from the wings of atomic density profile $n(r,t)$, and $e(r,t)$ from the equation of state.  Step B:  we derive  $j_p(r,t)$ and $j_e(r,t)$ based on the continuity equations, where the particle $s_p(r,t)$ and energy source terms $s_e(r,t)$ come from three-body physics. Step C: We linearly decompose the fluxes in terms of thermodynamic force ansatz $f_p$ and $f_e$, which, along with the Onsager relations, yields the transport properties of the sample.  }
\label{fig1-5}
\end{figure}

In this article we report on the development of a new scheme to extract thermodynamic fluxes and kinetic coefficients from the time-resolved $\mathit{in}$ $\mathit{situ}$ images of 2D atomic quantum gases, see Fig. \ref{fig1}. For our system the dynamics are driven by weak three-body recombination (3BR) near a Feshbach resonance \cite{RMPFeshReview}, which induces particle loss and heating so slowly ($\sim1$~s) that the evolution is essentially quasi-static. The extraction of transport properties is realized in three steps (see Fig. \ref{fig1-5}).  Step A: We analyze the $\mathit{in}$ $\mathit{situ}$ density profiles at different times $t$ and determine the evolution of the global temperature $T(t)$ and chemical potential $\mu_0(t)$ at the center of the trap, see Fig. 1c and 1d. From the equation of state measurement \cite{Strong2D} we obtain the energy density distribution $e(r,t)$, where $r$ is the distance from the trap center. Step B:  We determine the particle and energy fluxes based on the continuity equations and microscopic models for the particle loss and heating due to three-body processes.  Step C: By decomposing the fluxes in terms of a thermodynamic force ansatz we determine the essential kinetic coefficients, as described by equations (\ref{Fluxeqn}) and (\ref{Fluxeqn1}), including the thermoelectric Seebeck coefficient $\mathcal{S}$ and Lorenz number $\mathcal{L}$.

Three-body processes occur predominately in regions of high density and thus reduce the atomic density near the trap center, see Fig. \ref{fig1}b. Away from the center, remarkably, we observe an increasing density for $r>35~\mu$m, indicated by the green arrow (see Fig. \ref{fig1}b), which is a result of heating induced by 3BR. This observation suggests the need to describe the transport by two processes. While 3BR loss induces an inward particle flow to the trap center, it also deposits energy into the sample [24] and drives a thermoelectric current that flows outward; see Fig. \ref{fig1}e and \ref{fig1}f.  The total particle flux is a result of these two effects.

To quantify the particle flux $j_p(r,t)$, we model the 3BR loss as $s_p=-k n(r,t)^3$, and integrate the continuity equation, which yields

\begin{equation}
2\pi r j_{p}(r,t)=-\int_0^rdA[\partial_tn(r',t) + k n(r',t)^3],
\label{partflux}
\end{equation}
\noindent  where $dA=2\pi r'dr'$. The three-body loss coefficient $k$ is determined from the particle flux boundary condition, $j_p(r=\infty,t)=0$, see methods.

\begin{figure}[th!]
\includegraphics[height=5 in]{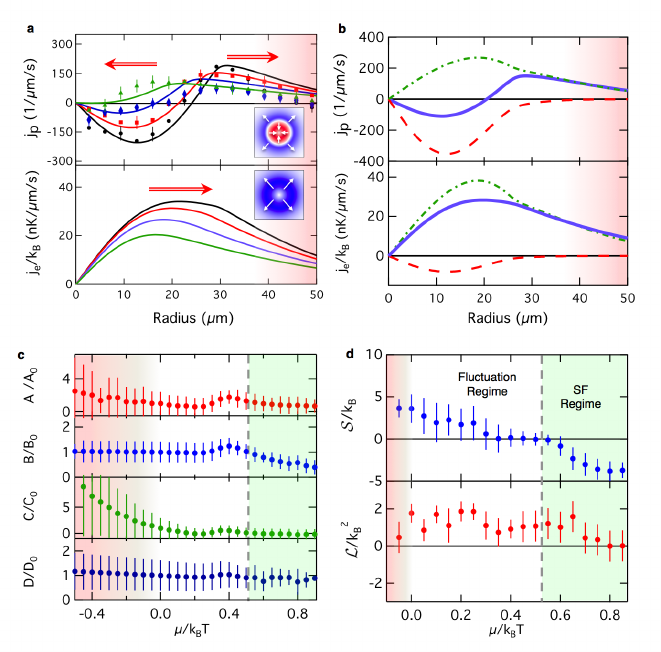}
\caption{\begin{bf}Extraction of particle and energy fluxes, and transport coefficients.\end{bf}  $\mathbf{a}$, the calculated particle (upper panel) and energy (lower panel) fluxes in the system. The data points are based on direct differentiation of the data at 10~ms (black dots), 50~ms (red squares), 110~ms (blue diamonds), and 190~ms (green triangles). Solid lines show the corresponding fluxes from parameterized density profiles. The red arrows indicate the direction of particle and energy flux. The insets illustrate the fluxes in the trap, where red(blue) indicates inward (outward) flow.   $\mathbf{b},$ The decompositions of $j_p$ (upper panel) and $j_e$ (lower panel) at $t$=60 ms in terms of the force ansatz $f_p$ (red dashed line) and $f_e$ (green dot-dashed line). $\mathbf{c},$ The extracted normalized transport matrix coefficients, where $A_0=A(\mu=0)$, $B_0=B(\mu=0)$... $\mathbf{d},$ Seebeck coefficient $\mathcal{S}$ and Lorenz number $\mathcal{L}$.  The green shaded area indicates the superfluid regime \cite{Strong2D}, and the red shading indicates the crossover from hydrodynamic to ballistic (collisionless) regime. The vertical dashed line indicated the superfluid transition.}
\label{fig4}
\end{figure}

The calculated particle flux $j_p(r,t)$, shown in Fig.~\ref{fig4}a, clearly suggests that there are two regions of transport. In the interior region, particles flow inward $j_p<0$, and, in the exterior region, the flow is outward $j_p>0$,  see inset of Fig.~\ref{fig4}a. The latter, as we will see, manifests the thermoelectric effect.

To extract the energy distribution $e(r,t)$, we verify and use the scale invariance of the equation of state $n(\mu,T)\lambda_{dB}^2=F(\tilde{\mu})$ \cite{CLHNature,Strong2D,Dalibard2DThermo} to parameterize the energy density $e(\mu,T)=\lambda_{dB}^{-2} k_B T \int^{\tilde{\mu}}F(x)dx$ \cite{Dalibard2DThermo}, where $\tilde{\mu}=\mu/k_BT$, $\lambda_{dB}=\sqrt{\frac{h^2}{2\pi m k_B T}}$ is the de Broglie wavelength, $m$ is the atomic mass, $k_B$ is Boltzmann's constant, $h$ is Planck's constant, and $F(x)$ is a generic function \cite{CLHNature}, see supplemental material. Together with the local density approximation $\mu(r,t)=\mu_0(t)-V(r)$, trap potential $V(r)$, and temperature $T(t)$, we obtain the energy density distributions $e(r,t)$.

To investigate the thermoelectric effect, we evaluate the energy flux $j_e(r,t)$ based on the continuity equation and a microscopic model to describe the heating due to 3BR. The energy source term is introduced as $s_e=-H s_p$, where $H$ is the energy gain per lost atom. Incorporating both the interaction and potential energy of an atom $E=e/n+V$, we have $H=(\bar{E}-E)+k_BT_h$, where in the parenthesis is the anti-evaporative heating \cite{Grimm3BR}, $\bar{E}$ is the mean energy of a trapped atom, and $k_BT_h$ accounts for additional energy transfer occurring in the loss process \cite{Grimm3BR,SheddingLight3BR}. To sum up, we determine the energy flux $j_e(r,t)$ as

\begin{align}
2\pi r j_e(r,t)=&-\int_0^rdA[\partial_te(r',t) +H k n(r',t)^3].
\label{enflux}
\end{align}

\noindent Based on the boundary condition for energy $j_e(r=\infty,t)=0$, we find $T_h=21\sim85$~nK, consistent with previous study on cesium \cite{Grimm3BR}. The derived energy flux, shown in Fig. \ref{fig4}, suggests that the dominance of outward energy flux $j_e>0$ for all $r$, a result of strong 3BR heating with radial temperature gradient $\nabla_r T<0$.

With $j_p(r,t)$ and  $j_e(r,t)$ determined we proceed onto Step C to extract the kinetic transport coefficients as shown in equations (1) and (2). In principle, one can extract the forces $f_p(r,t)$ and $f_e(r,t)$ by measuring $\nabla (\mu/T)$ and $\nabla(1/T)$, but these gradients are too small in our quasi-static process to be measured with sufficient precision. As an alternative, we devise a phenomenological approach to determine the functional forms of $f_p(r,t)$ and $f_e(r,t)$. To do this we identify that the ansatz $f_p\propto -r n^{\alpha_p}$ and $f_e\propto r n^{\alpha_e}$ can well describe the fluxes in different regimes. Here the exponents $\alpha_p$ and $\alpha_e$ are fitting parameters. The ansatz can be understood in the following: the leading factor $r$ reflects the restoring force  in a harmonic trapping potential, and the 3BR process suggests a dominating dependence on density $n$. Finally, the different signs come from the physical consideration that atom loss and heating result in $\nabla_r\mu>0$ and $\nabla_r T<0$, which lead to negative $f_p=-\nabla_r (\mu/T)<0$ and positive $f_e=\nabla_r(1/T)>0$.

With this ansatz we fit the fluxes according to

\begin{equation}
\left(\begin{array}{c} j_p(r,t) \\ j_e(r,t) \\ \end{array}\right)=\left(\begin{array}{cc}A(\tilde{\mu}) & B(\tilde{\mu}) \\ C(\tilde{\mu}) &D(\tilde{\mu}) \\ \end{array} \right)
\left(\begin{array}{c} -r n(r,t)^{\alpha_p}\\r n(r,t)^{\alpha_e}\ \\ \end{array}\right),
\label{Fluxeqn2}
\end{equation}

\noindent where the matrix elements $A$, $B$, $C$ and $D$ are constrained to depend only on the state of the system $\tilde{\mu}$. A complete survey of the fluxes between times 0~$\sim$~200 ms in 20~ms intervals shows that the above ansatz can very well capture the behavior of the fluxes with exponents $\alpha_p=3.2(3)$ and $\alpha_e=0.9(2)$. The different exponents suggest that the chemical potential gradient $\propto n^{3.2}$ dominates at small radii and a temperature gradient $\propto n^{0.9}$ dominates at large radii, which is consistent with our picture. An example of the decomposition and the extracted coefficients are given in Fig.~\ref{fig4} b and c (see supplemental material).

Comparing equations (\ref{Fluxeqn}), (\ref{Fluxeqn1}), and (\ref{Fluxeqn2}), we find that the Peltier coefficient $\mathcal{P}$, Seebeck coefficient (thermopower) $\mathcal{S}$ and the Lorenz number $\mathcal{L}$ are given by

\begin{equation}
\begin{aligned}
\mathcal{S}=&\frac{\mathcal{P}}{T}= \frac 1T\left( \frac {C}{A} - \tilde{\mu}\right)\\
 \mathcal{L} =& \frac C{T^2A}\left(\frac{D}{B}-\frac{C}{A}\right),
 \end{aligned}
\label{Seebeck}
\end{equation}

\noindent where we have used the Onsager relation to obtain $\mathcal{S}$ and $\mathcal{L}$. The result is shown in Fig.~\ref{fig4}d.

It is important to note that for the transport theory to apply, the gas must be in the hydrodynamic regime, with mean-free path $l$ much smaller than the cloud size and scattering rate faster that trap frequency. Given our cloud size of 50~$\mu$m, calculations show that for $\tilde{\mu}>0$ (fluctuation and superfluid regimes) our gas is hydrodynamic with $l=1\sim 5~\mu$m, and in the normal gas phase $\mu<0$, the system gradually enters the collision-less regime\cite{PetrovShlyapnikovPRA2001}.  The collisionless regime is marked by the red shaded area in  Fig.~\ref{fig4}, where the transport theory is expected to break down.

In the hydrodynamic regime, our result shows that $\mathcal{S}>0$ in the normal state (Fig. \ref{fig4}d), consistent with the notion that thermal atoms are the sole carriers of both the mass and the entropy in the normal phase. Approaching the superfluid transition, $\mathcal{S}$ decreases toward zero and becomes negative in the superfluid regime. The sign change of $\mathcal{S}$ occurs close to the superfluid transition point at $\tilde{\mu}=0.53(10)$ \cite{Strong2D}. This surprising result indicates that, in the absence of chemical potential gradient, a 2D Bose superfluid  moves against the heat flow (a.k.a. superfluid counterflow), namely, from the colder to the hotter side (see supplemental material). Superfluid counterflow is well known in He II as the origin of fountain effect. Beyond bosonic atoms, anomalous (non-zero and sign-changed) Seebeck coefficients were first discussed in the context of non-uniform superconductors \cite{GinsburgFiz14}, and were also observed recently in high-Tc material \cite{Taillefer}.

The Lorenz number $\mathcal{L}$ in the hydrodynamic regime is overall positive, consistent with the second law of thermodynamics, which demands $|\hat{\sigma}|=\sigma^2T^4\mathcal{L}\geq0$ \cite{MahanManyParticle}. In the superfluid regime, Lorenz number $\mathcal{L}$ gently reduces, but the large experimental uncertainties hinder our ability to determine its asymptotic behavior. Theoretically, thermoelectric properties are well studied for electron systems. Important results include the Mott formula for the thermopower $\mathcal{S}$ and the Wiedemann-Franz law for the Lorenz number $\mathcal{L}$ \cite{MottWFLaws}. For Bosons, Lorenz number $\mathcal{L}$ for a 2D thermal gas is expected to be a constant \cite{Mott2DTransport}, but there are no predictions across the fluctuation and superfluid regimes; calling for the need to develop the equivalence of the Mott and Wiedemann-Franz laws for bosonic systems.

In conclusion, our results comprise the first exploration into the particle and energy transports in a 2D Bose gas. In doing so we have presented a new method based on $\mathit{in}$ $\mathit{situ}$ imaging to unveil the intriguing thermoelectric properties of 2D gas in the fluctuation and superfluid regimes. Our approach can be extended by carefully mapping out the temperature and chemical potential gradients in order to completely characterize all kinetic coefficients independently. The transport properties of ultracold bosonic systems can potentially help understand thermoelectricity in condensed matter systems \cite{TEinUltracold}. Beyond condensed matter, transport measurement in quantum critical gases may also test the gauge-gravity duality conjecture \cite{QCBlackHoles,CriticalDynamicsSachdev}, as well as to identify new exotic phases of atomic quantum gases.

We would like to acknowledge Chih-Chun Chien, Charles Grenier, and Colin Parker for useful discussions and Logan Clark for careful reading of the manuscript.  This work is supported by NSF Grant No. PHY-0747907 and under ARO Grant No. W911NF0710576 with funds from the DARPA OLE Program.

\section{Methods}
Our experiment is performed based on a gas of  $\sim5\times 10^4$ cesium atoms at a temperature of 36~nK loaded into a two-dimensional optical trap with trap frequencies $(\omega_x,\omega_y,\omega_z)=2\pi\times(8,10,1900)$~Hz. For more details see references \cite{CLHNature,Strong2D}.  From the $\mathit{in}$ $\mathit{situ}$ measurement of the density profile $n(r,t)$ we can extract the temperature $T$ and chemical potential $\mu_0$ of the gas at the center of the cloud by fitting the tail of the density profile \cite{CLHNature,Dalibard2DThermo}. A linear magnetic field sweep over 200~ms is used to tune the scattering length to 22~nm.  We verify that this ramp does not cause any collective excitation. We then monitor the density profile $n(r,t)$ in 20~ms increments up to a hold time of $t=$200~ms.  This forms the basis of our analysis.

To verify that 3BR is the sole source that drives the dynamics, we start with a few-body collision loss model $\dot{n}=-k n^\alpha$, insert the measured 2D density $n(r,t)$, and integrate both sides of this equation. We find that the exponent  $\alpha=3$ unambiguously fits our data, and the loss coefficient is $k = 1.34(8)\times10^{-18}$~cm$^{-4}$s$^{-1}$. At much smaller scattering lengths ($<$10~nm), 3-body collisions are strongly suppressed and we observe negligible trap loss and heating based on the same experimental setup.
\bibliographystyle{naturemag}
\bibliography{AnomalousThermoelectricTransport}

\section{supplemental information}

\subsection{Equation of state and parameterization of thermodynamics}
From the $\mathit{in}$ $\mathit{ situ}$ images we determine the thermodynamic quantities as a function of time. We first extract $\mu_0$ and $T$ from the density tail and the result is fit to an evaporation model: (see Fig.~1c and 1d)
\begin{equation}
\begin{aligned}
T(t)=&T(0)e^{1-(1+8.5t)^{-1/2}}\\
  \mu_0(t)=&\frac{\mu_0(0)}{{1+1.8t}}.
 \end{aligned}
\label{TempandChem}
\end{equation}

\begin{figure}[t!]
\includegraphics[width=5 in]{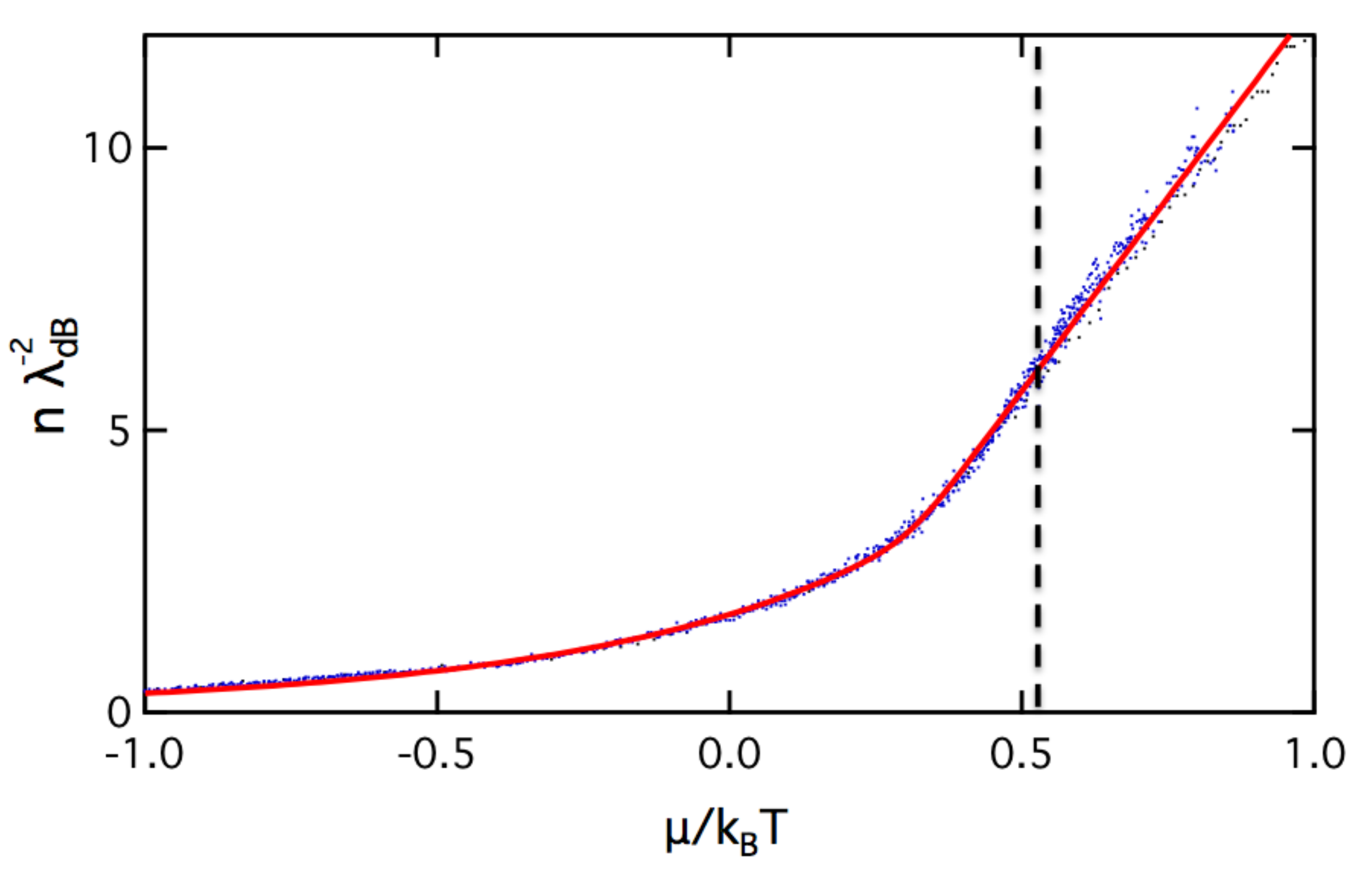}
\caption{\begin{bf} Phase space density of 2D Bose gas at $a=22$~nm.\end{bf} All of the experimental data points for hold times 0$\sim$200 ms in 20~ms steps are shown in dots and the fit based on our model is given as the solid red line.}
 \label{Sup1}
\end{figure}

\begin{figure}[t]
\includegraphics[width=4 in]{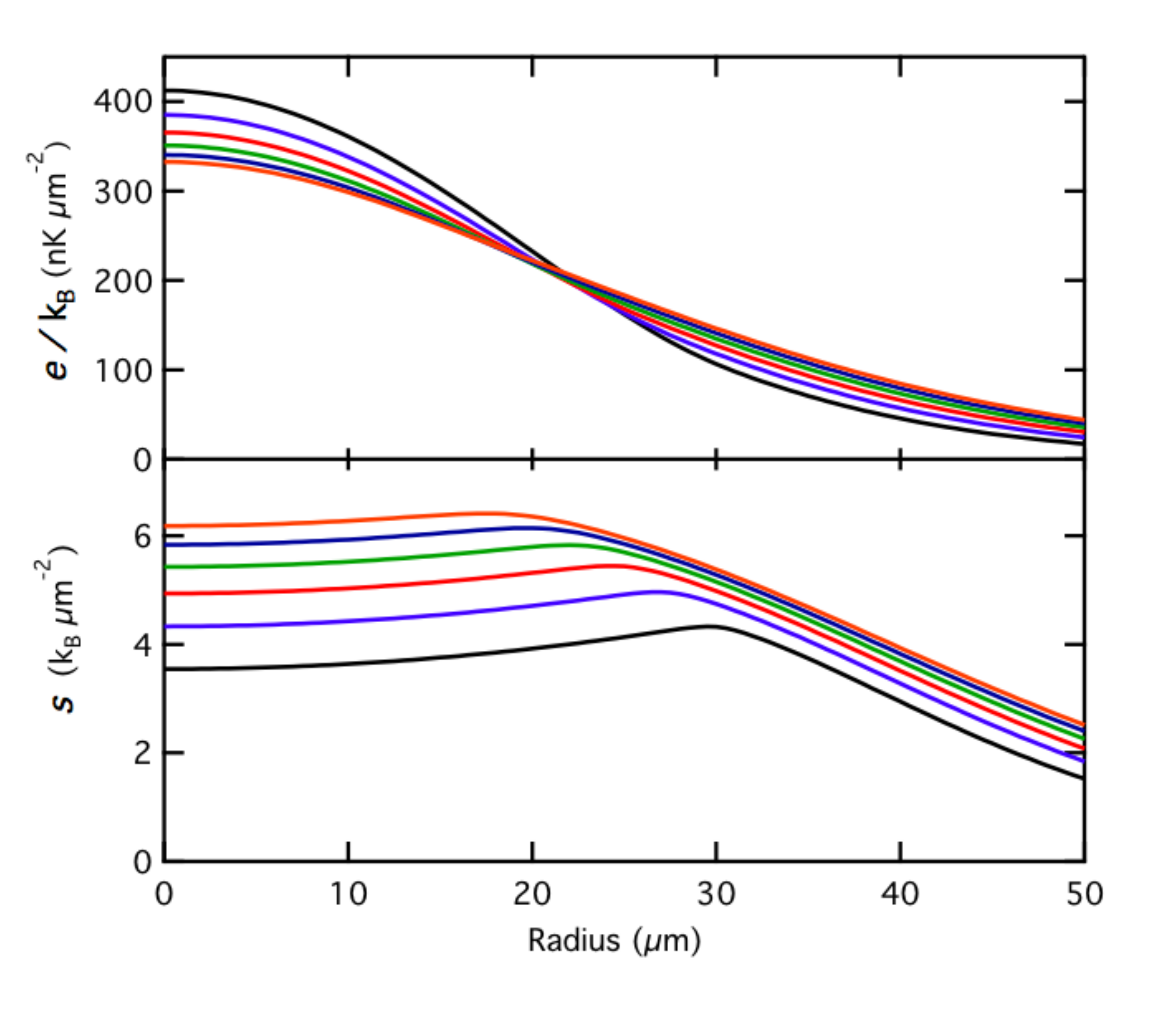}
\caption{\begin{bf} Evolution of energy density and entropy density\end{bf} The black curve represents the distribution at holding time $t$=0~ms and the other curves show the profiles with hold times in 40~ms intervals.}
 \label{Sup2}
\end{figure}

With the gas fully characterized for each time step we obtain the equation of state by plotting the phase space density $n(\mu,T) \lambda_{dB}^{2}=F(\mutilde)$ as a function of $\mutilde$, see Fig. \ref{Sup1}. The data points from all the measurements fall onto one curve, see Fig. \ref{Sup1}, which shows the scale invariance and confirm the quasi-static evolution of our 2D system. From the scale invariance we parameterize other thermodynamical quantities, for example, internal energy density $e$, entropy density $s$ and mean energy of a trapped atom $\bar{E}=<e/n+V>$,

\begin{eqnarray}
e(\mu,T)&=&\lambda_{dB}^{-2} k_B T \mathbf{F}(\tilde{\mu}) \label{scaleeqn} \\
s(\mu,T)&=&\lambda_{dB}^{-2}[2\mathbf{F}(\tilde{\mu})-\tilde{\mu}F(\tilde{\mu})] \nonumber\\
\bar{E}(\mu,T)&=&2k_B T \mathbb{F}(\tilde{\mu})/\mathbf{F}(\tilde{\mu}),\nonumber
\end{eqnarray}
\noindent where $\mathbf{F}(\tilde{\mu})= \int^{\tilde{\mu}}F(x)dx$ and $\mathbb{F}(\tilde{\mu})=\int^{\tilde{\mu}}\mathbf{F}(x)dx$.  The energy and entropy profiles of our gas are shown in Fig. \ref{Sup2} .

To determine the generic function $F(\tilde \mu)$ we model the isothermal compressibility with contributions from the normal and the superfluid component $\kappa = \frac{\partial n}{\partial \mu}= f_{n} \kappa_{n} + f_{s}\kappa_{s}$, where $f_{n(s)}$ and $\kappa_{n(s)}$ are the normal (superfluid) fraction and the associated compressibility, respectively. The superfluid compressibility $\kappa_{s}$ is given by the coupling constant \cite{Strong2D}. The normal gas compressibility is modeled as $\kappa_{n} = \exp\frac{\mu }{k_B T} + a \exp\frac{2\mu }{k_B T}$ and the thermal fraction as $f_{n}= 1/(1 + c \exp\frac{d\mu }{k_B T})$ and the superfluid fraction is $f_{s}=1-f_{n}$. By integrating $\kappa$ with respect to $\mu$ we obtain an analytic form of the density to fit the equation of state $n(\mu,T)\lambda_{dB}^2=F(\mutilde)$ with three fitting parameters $a$, $c$, and $d$.

\subsection{Decomposition of $j_p(r,t)$ and $j_e(r,t)$}

We linearly decompose the particle and energy fluxes based on the thermodynamic force ansatz $f_p\propto -r n^{\alpha_p}$ and $f_e\propto -r n^{\alpha_e}$. To determine the exponents we base on our observation that at large radii both $j_e(r,t)$ and $j_e(r,t)$ are dominated by the ``heating'' force $f_e$ and a fit to the fluxes at all times determines the exponent $\alpha_e=0.9(2)$.  With this knowledge we then decompose the fluxes at all radii in terms of $r n^{\alpha_p}$ and $r n^{\alpha_e}$, from which we find $\alpha_p=3.2(3)$.  For the data in Fig.~(3) we adopt the ansatz with $\alpha_p=3$ and $\alpha_e=1$ to decompose and determine the transport properties. We also attempt the decomposition with other combinations $(\alpha_p ,\alpha_p)=(3.2,0.9),(2,1),(4,1)...$ and observe only minor changes of the extracted transport properties. Our conclusion on the general behavior of the Seebeck and Lorenz number is insensitive to the above choices.

\subsection{Negative Seebeck coefficient and superfluid counterflow}

\begin{figure}[t]
\includegraphics[width=4 in]{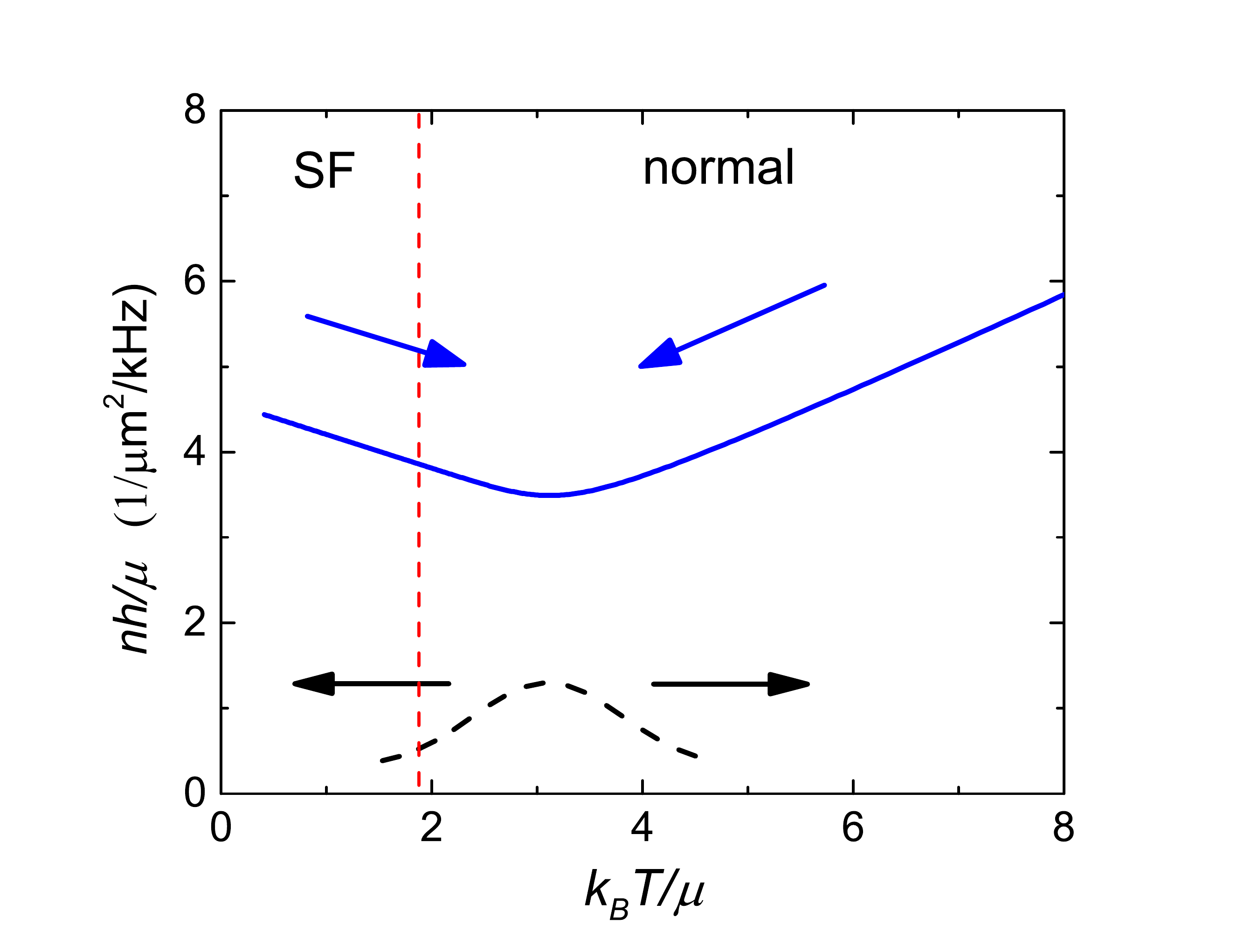}
\caption{\begin{bf} Illustration of thermoelectric current and Seebeck effect in the presence of temperature gradient\end{bf}. Based on the equation of state determined in Fig.~\ref{Sup1}, we evaluate the atomic density distribution (blue line) in a thought experiment where the chemical potential is constant and positive $\mu>0$, and the temperature is higher on the right side. Superfluid phase transition is indicated by the red line. The blue arrows show the direction of the diffusion current toward the lower density region. In a new equilibrium, a chemical potential barrier is formed (black line) to balance the diffusion flow (black arrows).}
\label{Sup3}
\end{figure}

Here we offer a thermodynamic picture to elucidate the observed anomalous behavior of Seebeck coefficient (thermopower) $\mathcal{S}<0$ in the superfluid regime. In a thought experiment, a 2D Bose gas is initially prepared and held at a constant chemical potential $\mu>0$ and with a temperature gradient. The expected density distribution (see Fig.~\ref{Sup3}) has a minimum near $k_BT/\mu\sim 3$. After releasing the atoms, they thermally diffuse toward the lower density region (blue arrows) and eventually reach a new equilibrium by forming a chemical potential barrier. The barrier results in a conductive current (black arrows) that balances the diffusive flow. Note that superfluid diffuses toward the hotter side (superfluid counterflow).

Thermopower $\mathcal{S}=-\nabla\mu/\nabla T|_{j_p=0}$ is defined as the amount of chemical potential gradient built up by the temperature gradient when the new equilibrium is reached. Where the two gradients have the same sign, as is in the superfluid phase, we obtain $\mathcal{S}<0$. The negative thermopower can thus be attributed to the counterflow or the negative dilatation constant $\eta=\partial n/\partial T|_{\mu}<0$.

Remarkably, superfluid counterflow cannot be responsible for the heat flow, which moves toward the colder side (second law of thermodynamics). Plausible understanding of entropy transport in superfluids with negative thermopower may involve a two-fluid model with a thermal component that follows the heat flow, or the creation of vortices as the entropy carriers.

\end{document}